\documentstyle[preprint,aps]{revtex}
\preprint {IMSc/96/08/26, 
IP-BBSR-96/44}

\begin{document}

\newcommand\beq{\begin{equation}}
\newcommand\eeq{\end{equation}}
\newcommand\bea{\begin{eqnarray}}
\newcommand\eea{\end{eqnarray}}

\draft

\title{Novel correlations in two dimensions: Two-body problem}

\author{R. K. Bhaduri$^a$, Avinash Khare$^b$, J. Law$^c$, M. V. N. 
Murthy$^a$\thanks{Permanent Address: The Institute of Mathematical Sciences,
Madras 600 113, India} and Diptiman Sen$^d$}

\address
{$^a$Department of Physics and Astronomy, McMaster University,\\ 
Hamilton L8S 4M1, Ontario, Canada \\
$^b$ Institute of Physics, Bhubaneswar 751 005, India\\
$^c$ Department of physics, University of Guelph, Guelph N1G 
2W1, Ontario, Canada\\
$^d$ Centre for Theoretical Studies, Indian Institute of 
Science,\\ Bangalore 560 012, India } 
\date{\today}
\maketitle
\begin{abstract}
We discuss a many-body Hamiltonian with two- and three-body interactions 
in two dimensions introduced recently by Murthy, Bhaduri and Sen. Apart 
from an analysis of some exact solutions in the many-body system, we 
analyze in detail the two-body problem which is completely solvable. We show 
that the solution of the two-body problem reduces to solving a known 
differential equation due to Heun. We show that the two-body spectrum 
becomes remarkably simple for large interaction strength and the level 
structure resembles that of the Landau Levels. We also clarify the 
"ultraviolet" regularization which is needed to define an inverse-square 
potential properly and discuss its implications for our model.
\end{abstract} 
\vskip 1 true cm

\narrowtext
\section{Introduction}

Recently, there has been a revival of interest in the area of exactly
solvable models in one and higher dimensions.  A celebrated example of a
solvable many-body system is the well-known Calogero-Sutherland model
(CSM) in one dimension \cite{C69,S71,S72}. The model has found wide
application in areas as diverse as quantum chaos and Fractional
Statistics. The particles in the CSM are confined in a one-body oscillator
potential or on the rim of a circle, and interact with each other through
a two-body potential which varies as the inverse-square of the distance
between particles. The CSM and its variants in one dimension, like the
Haldane-Shastry model for spin chains \cite{H88}, have provided the
paradigms to analyze more complicated interacting systems. A
characteristic feature of the CSM is the structure of the highly
correlated wave function. The correlations are built into the exact wave
function through a Jastrow factor $(x_i -x_j)^{\lambda}
|x_i-x_j|^{\alpha}$ for any pair of particles denoted by $i,j$. The
exponents on the correlator are related to the strength of the
inverse-square interaction. Notice that this factor is antisymmetric
(symmetric) in particle labels for $\lambda=1(0)$ and vanishes as the two
particles approach each other. A generalization of this in two dimensions
is to be found in Laughlin's trial wave function \cite{L83} where the
correlations are built in through the factor $(z_i - z_j)$, where $z_i$
are the particle coordinates in complex notation. The corresponding
Hamiltonian for which the Laughlin wave function is an exact ground state
has not been analyzed to the same degree of detail as the CSM. It is known
that it is the ground state for a Hamiltonian describing spin polarized
electrons in the lowest Landau level with a short-range repulsive
interaction \cite{TK85}. It is also known that such correlations are
present in the exact ground state of a spin Hamiltonian\cite{L89} in
two-dimensions.  The anyon Hamiltonian\cite{LM79} in two-space dimensions
is another example where the Jastrow correlation appears\cite{WU84}. While
the two-anyon problem is exactly solvable, the many-anyon problem is not.
For a system of anyons confined in an oscillator potential many exact
solutions and their properties are known but, unlike the CSM in
one-dimension, the analytical solution of the full many-body problem is
not tractable\cite{exact}. It is therefore of great interest to find
models analogous to the CSM in higher dimensions. 

In a recent paper\cite{MBS96}, three of us proposed a model in two-space 
dimensions with nontrivial two- and three-body interactions which could 
be solved exactly for the ground states and some excited states. It 
betrayed some similarity to both CSM in one-dimension and the anyonic 
model in two-dimensions through the spectrum. The model was devised by noting 
that in two dimensions there exists another form of the pair 
correlator with which a Jastrow-type many-body wave function may be 
constructed, namely  
\beq
X_{ij} ~=~ x_i y_j ~-~ x_j y_i ~.  
\label{xij} 
\eeq
The correlation is by definition antisymmetric and goes to zero as two
particles approach each other. In addition, it introduces zeros in the
wave function whenever the relative angle between the two particles goes
to zero or $\pi$. The difference with the Jastrow-Laughlin form is also
significant; $X_{ij}$ in (\ref{xij}) is a pseudo-scalar. Unlike the Laughlin 
type of correlation, it does not impart any angular momentum to the two
dimensional wave function . One important drawback of this correlation is
that it is not translationally invariant unless the radial degrees of
freedom is frozen. The model Hamiltonian has solutions which have this
correlation built in.  Intuitively the correlation can be understood easily
by imagining objects with associated "arrows". The arrows cannot be
oriented either parallel or anti-parallel to each other. 
The model has some interesting features and it would be of great
interest to find physical systems which incorporate these features. 

In this paper we elaborate on our earlier results \cite{MBS96} and 
present several 
new results. In Sect. II, we discuss the many-body Hamiltonian and display
some of the exact solutions and their structure. The similarities between
the spectrum of these exact solutions and the spectrum of CSM is quite
remarkable. Further, when projected on to a circle the model reduces to a
variant of the trigonometric Sutherland model. In this limit
the model also has translational invariance.  In Sect. III, we discuss
the two-body problem in detail and show that the solutions of the two-body
problem are described by the Heun equation. In particular, the spectrum 
becomes very simple for large values of the interaction strength. The singular
interaction discussed in this paper requires a careful treatment 
in the region near $X_{ij} =0$; this is discussed in the Appendix \cite{S58}.
Sect. IV contains a discussion and summary. 

\section{Many-body Hamiltonian and some exact solutions}

For the sake of completeness we recall first the Hamiltonian and some 
of its properties proposed earlier\cite{MBS96}. We also clarify 
some points which were not made explicitly clear in the earlier 
paper. 
The $N$-particle Hamiltonian  which we consider is given by 
\beq
H~=~ -{\hbar^2\over {2m}} ~\sum_{i=1}^{N} {\vec \nabla}^2_i + {{m\omega^2}
\over 2} ~\sum_{i=1}^{N} {\vec r}_i^2 + \frac{\hbar^2}{2m} g_1 ~
\sum_{\stackrel{i,j}{ (i\ne j)}}^{N} \frac{{\vec r}_j^2}{X_{ij}^2}
+ \frac{\hbar^2}{2m} g_2 ~\sum_{\stackrel{i,j,k}{ (i\ne j\ne k)}}^{N} 
\frac{{\vec r}_j \cdot {\vec r}_k}{X_{ij}X_{ik}} ~,    
\eeq
where $X_{ij}$ is given by (\ref{xij}); $g_1$ and $g_2$ are
dimensionless coupling strengths of the two- and three-body interactions
respectively. While $g_1$ and $g_2$ can be independent of each 
other in general, for the type of solutions involving the correlator in 
(\ref{xij}) they are not. We will specify their relationship shortly. The 
particles are confined in a one-body oscillator
confinement potential. The Hamiltonian is rotationally
invariant and manifestly symmetric in all particle indices. As in the 
CSM, we may scale
away the mass $m$ and oscillator frequency $\omega$ by scaling all distances
$\vec r_i \rightarrow \sqrt{m\omega/\hbar}~ \vec r_i$, and measuring the
energy in units of $\hbar \omega$. This is done by setting
$\hbar =m=\omega =1$. In these units the Hamiltonian is given by
\beq
H~=~ -{1\over 2} ~\sum_{i=1}^{N} {\vec \nabla}^2_i +{1\over 2} ~\sum_{i=1}^{N} 
{\vec r}_i^2 + \frac{g_1}{2} ~\sum_{\stackrel{i,j}{(i\ne j)}}^N 
\frac{{\vec r}_j^2}{X_{ij}^2} + \frac{g_2}{2} ~\sum_{\stackrel{i,j,k}{(i\ne 
j\ne k)}}^N \frac{{\vec r}_j \cdot {\vec r}_k}{X_{ij}X_{ik}} ~. 
\label{ham2} 
\eeq
Note that the total angular momentum operator
\beq L ~=~ \sum_i ~(x_i p_{y_i} - y_i p_{x_i}) 
\label{angmom} 
\eeq
commutes with the Hamiltonian since it is rotationally invariant, and may 
therefore be used to label the states. The Hamiltonian is invariant under 
parity $x\rightarrow -x$ and $y\rightarrow y$. In addition, for any $i$, 
the Hamiltonian is invariant under the transformation $\vec r_i
\rightarrow -\vec r_i$ and $\vec r_k \rightarrow \vec r_k$ for all $k \neq i$. 
This $D_{2N}$ invariance is special to this system, and we are not aware of 
any other interacting
many-body Hamiltonian which has this symmetry. The consequences of this will 
be discussed explicitly in the two-body problem where this is related 
to the supersymmetric properties of the system. 

We will consider both bosonic and fermionic systems governed by the Hamiltonian
(\ref{ham2}), i.e., wave functions which are totally symmetric and 
antisymmetric respectively. It will turn out that certain calculations (for 
example, in the two-body problem) simplify if we do not impose any symmetry to
begin with.

\subsection{The exact bosonic ground state}

We first obtain the exact bosonic ground state of this Hamiltonian. As an 
ansatz for the ground state wave function, consider a solution of the form 
\beq
\Psi_0 (x_i,y_i)~=~ \prod_{i<j}^N ~|X_{ij}|^{g}~
\exp ~(-{1\over 2} \sum_{i=1}^N {\vec r}_i^{~2}) ~.    
\label{ansatz}
\eeq
Clearly $\Psi_0$ correctly incorporates the 
behavior of the wave function in the asymptotic region $|\vec r_i| 
\rightarrow \infty$, and $\Psi_0$ is regular for $g \ge 0$. In general
we insist that our solutions have this asymptotic form; the conditions under
which this is valid will be specified later. The eigenvalue equation now 
takes the form 
\beq
H \Psi_0 ~=~[\frac{1}{2} (g_1-g(g-1)) \sum_{\stackrel{i,j}{(i\ne j)}}^{N} 
\frac{{\vec r}_j^2}{X_{ij}^2} + \frac{1}{2} (g_2-g^2) \sum_{\stackrel{i,j,k} 
{(i\ne j\ne k)}}^{N} \frac{{\vec r}_j \cdot {\vec r}_k}{X_{ij}X_{ik}} + 
gN(N-1) +N] \Psi_0 ~. 
\eeq
Therefore $\Psi_0$ is the exact many-body ground state for an arbitrary 
number of particles of the Hamiltonian if 
\beq
g_1 ~=~ g(g-1) \quad {\rm and} \quad g_2 ~=~ g^2 ~.
\label{g12}
\eeq
Since $g \ge 0$, we have $g_1 \ge -1/4$ and $g_2$ is positive 
definite. Note that the range of $g_1$ is identical to the one obtained in 
the CSM. The ground state energy is now given by 
\beq 
E_0 ~=~ N ~+~ gN(N-1) ~.   
\label{egs}
\eeq
Note that this has exactly the form of the ground state energy of the CSM. 

Since $g$ determines both $g_1$ and $g_2$ uniquely, we will regard $g$ as 
the fundamental parameter of the Hamiltonian which determines the strength of 
the interaction. In other words, we demand that the ground state should be 
given by (\ref{ansatz}), and we {\it define} the Hamiltonian to ensure that.
It turns out that such a definition requires some special care in the 
vicinity of $X_{ij} =0$ (called the ultraviolet region below). The Appendix 
will discuss this for the case of two particles. Note that in the two-particle 
case, the Hamiltonian only contains the parameter $g_1 = g(g-1)$ and not $g_2$.
As a result, for every value of $g_1$ in the range $-1/4 <g_1 < 0$,
the bosonic ground state energy as given by $E_0 = 2 +2g$ has two possible
values; these two possibilities correspond to different potentials in the
ultraviolet region. This is somewhat unusual but it is not uncommon for 
singular potentials. The same thing also happens in the CSM even for the 
$N$-body problem; see for example Ref. \cite{MS94}. We discuss this issue
in detail in the Appendix where we show that the ultraviolet
regularization is determined by the parameter $g$ rather than by $g_1$.

We emphasize that our objective here is not to find the general solutions
for arbitrary $g_1$ and $g_2$, but to find a Hamiltonian whose solutions
have the novel correlation in Eq. (\ref{xij}) built in. In general, if
$g_1$ and $g_2$ are independent, the Hamiltonian will have a ground state
different from the one given above. Our procedure is
therefore similar to the many-anyon problem where also there are two- and
three-body interactions, but the strengths are related to a single
parameter.  With the form of $g_1$ and $g_2$ given in (\ref{g12}),
the solution found above is indeed the lowest energy state. 

A neat way of proving that we have indeed obtained the ground state
can be given using the method of operators \cite{P92}. To this end, define 
the operators
\bea 
Q_{x_i} ~&=&~ p_{x_i} ~-~ ix_i ~+~ i~g ~\sum_{j(j\ne i)}
\frac{y_j}{X_{ij}} ~, \nonumber \\
Q_{y_i} ~&=&~ p_{y_i} ~-~ iy_i ~-~ i~g ~\sum_{j(j\ne i)}
\frac{x_j}{X_{ij}} ~, \nonumber \\
Q_{x_i}^{\dag} ~&=&~ p_{x_i} ~+~ ix_i ~-~ i~g ~\sum_{j(j\ne i)}
\frac{y_j}{X_{ij}} ~, \nonumber \\
Q_{y_i}^{\dag} ~&=&~ p_{y_i} ~+~ iy_i ~+~ i~g ~\sum_{j(j\ne i)}
\frac{x_j}{X_{ij}} ~. 
\eea
It is easy to see that the $Q$'s annihilate the ground state in Eq. 
(\ref{ansatz}), $$Q_{x_i}\Psi_0 = 0 \quad {\rm and} \quad Q_{y_i}\Psi_0 =0.$$ 
The Hamiltonian can now be recast in terms of these operators as 
\beq
\frac{1}{2} ~\sum_i ~\left[ Q_{x_i}^{\dag} Q_{x_i} + Q_{y_i}^{\dag} Q_{y_i} 
\right] ~=~ H ~-~ E_0 ~, 
\eeq
where $E_0$ is given by Eq. (\ref{egs}). Clearly the operator on the left hand 
side is positive definite and annihilates the ground state wave function given 
by Eq. (\ref{ansatz}). Therefore $E_0$ must be the minimum energy that an 
eigenstate can have. 

As we remarked earlier, the ground state of the Hamiltonian is bosonic. 
The ground state of the Hamiltonian for a fermionic system is not easy to 
determine analytically (for $g > 0$). The problem here is analogous to a 
similar problem in the many-anyon Hamiltonian \cite{KM91,anyon}. In Sect. III, 
we will determine the fermionic ground state energy for two particles both 
numerically and to first order in $g$ using perturbation theory near $g=0$ 
and show that it has quite unusual behavior.

\subsection{Spectrum of excited states}

While we have not been able to find the complete excited state spectrum of 
the model, the eigenvalue equation for a general excited state may be 
obtained as follows. From the asymptotic properties of the solutions of
the Hamiltonian in Eq. (\ref{ham2}), it is clear that $\Psi$ has the
general structure
\beq
\Psi (x_i,y_i) ~=~ \Psi_0 (x_i,y_i) ~\Phi(x_i,y_i) ~,
\eeq
where $\Psi_0$ is the ground state wave function. Obviously if $\Phi$ is 
a constant we recover the ground state. In general $\Phi$ satisfies the 
eigenvalue equation 
\beq
[~-{1\over 2} ~\sum_{i=1}^{N}{\vec \nabla}_i^2 ~+~\sum_{i=1}^{N} {\vec r}_i 
\cdot {\vec \nabla}_i ~+~ g ~\sum_{\stackrel{i,j}{(i\ne j)}} \frac{1}{X_{ij}} 
(x_j \frac{\partial}{\partial y_i}-y_j \frac{\partial}{\partial x_i})~] ~\Phi ~
=~ (E - E_0) ~\Phi ~. 
\label{hamphi}
\eeq
It is interesting to note that while $g_1$ is zero both at $g=0$ and $1$, the 
term containing $g$ in the above expression is zero only when $g=0$. This 
is so because of the boundary condition that the wave 
functions must vanish as $|X_{ij}|^g$ for nonzero $g$. 

We first discuss the exact solutions of the above differential equation. 
This is easily done by defining the complex coordinates 
\beq 
z ~=~ x+iy \quad {\rm and} \quad z^* ~=~ x-iy ~,
\eeq 
and their partial derivatives
\bea
\partial ~=~ \partial/\partial z ~=~ \frac{1}{2} ~(\partial /\partial x -i 
\partial/\partial y)~, \nonumber \\
\partial^* ~=~ \partial/\partial z^* ~=~ \frac{1}{2} ~(\partial /\partial x +i 
\partial/\partial y) ~.
\eea
In these coordinates, the differential equation for $\Phi$ reduces to
${\tilde H} \Phi = (E-E_0) \Phi$, where 
\beq
{\tilde H} ~=~ -2\sum_i \partial_i\partial_i^* ~+~ \sum_i (z_i\partial_i 
+z_i^*\partial_i^*) ~+~ 2g\sum_{\stackrel{i,j}{(i\ne 
j)}}\frac{z_j\partial_i-z_j^*\partial_i^*}{z_i z_j^*-z_j z_i^*} ~.
\label{hcomplex}
\eeq
In addition, $\Phi$ is an eigenstate of the total angular momentum operator,
$L\Phi = l\Phi$. We can now classify some exact solution according to their 
angular momentum. 

\noindent
(a) $l=0$ solutions: Define an auxiliary parameter
$$ t ~=~ \sum_i z_i z_i^* ~,$$
and let $\Phi = \Phi(t)$. This has zero total angular momentum. 
The differential equation for $\Phi$ reduces to
\beq
t\frac{d^2 \Phi}{dt^2} +(b-t)\frac{d\Phi}{dt} -a\Phi ~=~ 0 ~,
\eeq
where $b=E_0$ and $a=(E_0-E)/2$; $E_0$ is the energy of the ground state. The 
allowed solutions are the regular confluent hypergeometric functions \cite{LL}
\beq
\Phi(t) ~=~ M(a,b,t) ~.
\eeq
Normalizability imposes the restriction $ a= -n_r $, where $n_r$ is 
a positive integer; then $\Phi(t)$ is a polynomial of degree $n_r$ (the 
subscript 'r' denotes radial excitations as discussed later). The 
corresponding eigenvalues are
\beq
E~=~ E_0 + 2n_r ~.
\eeq
This class of solutions was discussed before in \cite{MBS96}. 

\noindent
(b) $l > 0$ solutions: Let 
$$ t_z ~=~ \sum_i z_i^2 ~,$$
and let $\Phi = \Phi(t_z)$. 
The total angular momentum is not zero. All the mixed 
derivative terms in Eq. (\ref{hcomplex}) drop out, and we get the 
differential equation
\beq
2t_z\frac{d\Phi}{dt_z} ~=~ (E-E_0)\Phi.
\eeq
This is the well known Euler equation whose solutions are just monomials 
in $t_z$. The solution is given by
\beq
\Phi(t_z) ~=~ t^m_z ~,
\eeq
and the total angular momentum is $l=2m$. The eigenvalues are
\beq 
E ~=~ E_0 +2m ~=~ E_0 +l ~.
\eeq

\noindent
(c) $l < 0$ exact solutions: Let
$$ t_{z^*}=\sum_i (z_i^*)^2 ~,$$
and let $\Phi = \Phi(t_{z^*})$. Once again the 
differential equation for $\Phi$ reduces to
\beq
2t_{z^*}\frac{d\Phi}{dt_{z^*}} ~=~ (E-E_0)\Phi ~.
\eeq
This is similar to the previous case. The solution is given by
\beq
\Phi(t_{z^*}) ~=~ t_{z^*}^m ~,
\eeq
and the total angular momentum is $l=-2m$. The eigenvalues are 
\beq 
E ~=~ E_0 +2m ~=~ E_0 -l ~.
\eeq

\noindent
(d) Tower of excited states: One can now combine solutions of a given $l$ in 
cases (b) or (c) with the solutions in (a), and get a new class of excited 
states. Let us define
\beq
\Phi (z_i,z_i^*) ~=~ \Phi_1(t) \Phi_2(t_z) ~,
\eeq
where $\Phi_1$ is the solution with $l=0$, $\Phi_2$ is the solution 
with $l> 0$, and $t$ and $t_z$ have been defined before. The 
differential equation for $\Phi$ is again a confluent hypergeometric 
equation given by 
\beq
t\frac{d^2 \Phi}{dt^2} +(b-t)\frac{d\Phi}{dt} -a\Phi ~=~ 0 ~,
\eeq
where $b=E_0+2m$ and $a=(E_0+2m-E)/2$. The energy eigenvalues are then 
given by
\beq
E ~=~ E_0 + 2n_r + 2m ~=~ E_0 +2n_r +l ~.
\eeq
One may repeat the procedure to obtain exact solutions for a tower of 
excited states with $l<0$ solutions. As we shall see below, the 
existence of the tower is a general result applicable to all excited 
states of which the exact solutions shown above form a subset. We notice 
that these solutions bear a remarkable resemblance to the many-anyon 
system where a similar structure exists for the known class of 
exact solutions \cite{exact}. 

\noindent
(e) A general class of excited states: One can combine the solutions of all the 
three classes (a), (b) and (c) to obtain an even more general class of 
solutions. Consider the polynomial
\beq 
P(n_1,n_2,n_3) ~=~ t^{n_1}~ t_z^{n_2}~ t_{z^*}^{n_3} ~, 
\eeq
where the $n_i$ are non-negative integers. Using the form in (\ref{hcomplex}), 
one can show that 
\bea
{\tilde H} P(n_1,n_2,n_3) ~=~ &2& (n_1 + n_2 + n_3) P(n_1,n_2,n_3) ~-~ 8 n_2 
n_3 P(n_1+1,n_2-1,n_3-1) \nonumber \\
&-&~ 2 n_1 [n_1 + 2 n_2 + 2 n_3 + g N (N-1)] P(n_1-1,n_2,n_3) ~.
\eea
Using this one can show that there is an exact polynomial solution, whose
highest degree term is $P(n_1,n_2,n_3)$. The energy of this solution is
\beq 
E ~=~ E_0 ~+~ 2 (n_1 + n_2 + n_3) ~,
\eeq
and the angular momentum is $l = 2 (n_2 - n_3)$.

While there may be more exact solutions, we do not know of a simple way of 
solving for them. We can however glean some general features as follows. 
The coordinates $(x_i , y_i)$ can be separated into one `radial' coordinate 
$t = \sum_i {\vec r}_i^2$ as above and $2N-1$ `angular' coordinates 
collectively 
denoted by $\Omega_i$(say). Then, the Eq. (\ref{hamphi}) can be expressed as
\beq
t {{\partial^2 \Phi} \over {\partial t^2}} ~+~ (E_0 - t) {{\partial \Phi} 
\over {\partial t}} ~-~ \frac{1}{t} ~{\cal L} ~\Phi ~+~ \frac{1}{2} (E-E_0)~ 
\Phi ~=~ 0 ~,
\eeq
where ${\cal L} = {\cal D}_2 + g {\cal D}_1$, and ${\cal D}_n$ is an 
$n^{th}$-order differential operator which only acts on functions of the angles
$\Omega_i$. In particular, ${\cal D}_2$ is the Laplacian on a sphere of 
dimension $2N-1$. Next we note that the $\Phi$ can be factorized in the 
form 
\beq 
\Phi (x_i , y_i) ~=~ R(t) ~Y(\Omega_i) ~,
\eeq
where $Y$, generalized spherical harmonic defined on the $2N-1$ 
dimensional sphere $S^{2N-1}$, 
satisfies the eigenvalue equation ${\cal L} Y = \lambda Y$. (This is the hard
part of the spectral problem, to find the eigenvalues $\lambda$). We now define
\beq
\mu ~=~ {\sqrt {(E_0 -1 )^2 + 4 g \lambda}} ~-~ (E_0 -1) ~.
\eeq
Further if we write $R(t) = t^{\mu /2} {\tilde R} (t)$, then $\tilde R$ 
satisfies a confluent hypergeometric equation 
\beq
t \frac{d^2 {\tilde R}}{dt^2} ~+~ (b -t) \frac{d{\tilde R}}{dt} ~-~ a 
{\tilde R} ~=~ 0 ~. 
\eeq
where $b= E_0 + \mu$ and $a=(E_0+ \mu -E)/2$. The admissible solutions are 
the regular confluent hypergeometric functions, ${\tilde R} (t)=M(a,b,t)$. 
Normalizability imposes the restriction $a=-n_r$, where $n_r$ is a
positive integer. Then ${\tilde R} (t)$ is a polynomial of degree $n_r$, and it
has $n_r$ nodes. The energy of this state is given by $E = E_0 + \mu + 2n_r$. 
We see that for a given value of $\mu$, there is an infinite tower of energy 
eigenvalues separated by a spacing of $2$. As remarked earlier, this is 
reminiscent of what happens 
in the case of anyons. The tower structure and the angular momentum 
are useful in organizing a numerical or analytical study of the energy 
spectrum. Since the radial quantum number $n_r$ and the angular momentum $l$ 
are integers, they cannot change as the parameter $g$ is varied continuously. 

\subsection{Relation to Sutherland model}

It may be of interest to note that the model reduces to a variant of the 
Sutherland model\cite{S71} in one dimension. In this limit, therefore the 
model is exactly solvable.  Restricting the particles to 
move along the perimeter of a unit circle in the Hamiltonian (\ref{ham2}) 
without the confinement potential, we get 
\beq
H~=~ -{1\over 2} \sum_{i=1}^{N}{\partial^2\over \partial \theta_i^2} + 
\frac{g_1}{2} \sum_{\stackrel{i,j}{ (i\ne j)}}^{N} \frac{1}{\sin^2(\theta_i 
-\theta_j)} + \frac{g_2}{2} \sum_{\stackrel{i,j,k}{ (i\ne 
j\ne k})}^{N} [1+\cot(\theta_i-\theta_j)\cot(\theta_i -\theta_k)] ~, 
\label{ham5} 
\eeq
since $X_{ij}=-\sin (\theta_i -\theta_j)$ now. Using the identity
\beq
\sum_{\stackrel{i,j,k}{ (i\ne j\ne k)}}^{N} 
\cot(\theta_i-\theta_j)\cot(\theta_i-\theta_k) ~=~ -~ \frac{N(N-1)(N-2)}{3} ~,
\eeq
we immediately recover an analog of the trigonometric Sutherland model, but 
shifted by the constant $g_2~ N(N-1)(N-2)/3$. Note, however, that the  
potential in (\ref{ham5}) depends on the function $\sin(\theta_i 
-\theta_j)$,  rather than the chord-length which is proportional to 
$\sin[(\theta_i -\theta_j)/2]$ .  Interestingly the wave function has 
twice the periodicity of the Sutherland model solutions- the wave 
function vanishes whenever the particles are at diametrically opposite 
points on a circle or at the same point. 

\section{The two-body problem: Complete solution}

While we have not been able to solve the many-body problem completely, the 
two-body problem in our model is exactly solvable. We demonstrate this by 
going 
over to the hyperspherical formalism first proposed in two dimensions by 
Kilpatrick and Larsen \cite{KL87}(see also \cite{KM91}). We 
discuss some of the properties of
the two-body spectrum. We also explicitly show that the two-body problem is 
integrable. It is important to note that the two-particle interaction is
sufficiently singular that a careful treatment is required in order
to define the problem completely consistently; this is described in the 
Appendix.

The two-body Hamiltonian is given by
\beq
H ~=~ -\frac{1}{2}[{\vec \nabla}_1^2 +{\vec \nabla}_2^2] ~+~ \frac{1}{2}
[{\vec r}_1^2+{\vec r}_2^2] ~ +~ \frac{g_1}{2}
\frac{{\vec r}_1^2+{\vec r}_2^2}{X^2} ~,
\eeq
where $X =x_1 y_2 - x_2 y_1$. The two-body 
problem is best solved in the hyperspherical coordinate system which allows
a parameterization of the coordinates ${\vec r}_1, {\vec r}_2$ in terms of three 
angles and one length, $(R,\theta,\phi,\psi)$ as follows:
\bea 
x_1 + i y_1  ~&=&~ R~ (\cos \theta ~\cos \phi  - 
i\sin \theta ~\sin \phi) ~\exp (i\psi) ~, \nonumber \\  
x_2 + i y_2  ~&=&~ R ~(\cos \theta ~\sin(\phi) + 
i\sin \theta ~\cos \phi) ~\exp (i\psi) ~.
\eea
We may regard $(R,\theta,\phi)$ as the body-fixed coordinates which are 
transformed to the space-fixed system by an overall rotation of $\psi$. 
For a fixed $R$, these coordinates define a sphere in four-dimensions within
the following intervals:
\bea
-\pi/4 \le & \theta & \le \pi/4 ~, \nonumber \\
-\pi/2 \le & \phi & \le \pi/2 ~, \nonumber \\
-\pi~~ \le & \psi & \le \pi ~.
\eea
Exchange of two particles is achieved by
\bea
\theta & & \rightarrow -~ \theta ~, \nonumber \\
{\rm and} \quad \phi & & \rightarrow \pi/2 - \phi ~, \quad \psi \rightarrow 
\psi \quad {\rm if} \quad \phi > 0 ~, \nonumber \\
{\rm and} \quad \phi & & \rightarrow - \pi/2 - \phi ~, \quad \psi \rightarrow 
\pi + \psi \quad {\rm if} \quad \phi < 0 ~.
\label{exchange}
\eea
With this choice of coordinates, the radial coordinate becomes
\beq 
R^2 ~=~ r_1^2+r_2^2 
\eeq
which is the radius of the sphere in four dimensions. Also, 
\beq
X ~=~ x_1 y_2 -x_2 y_1 = R^2 \sin (2\theta) /2 ~.
\eeq
Notice that $X$ depends only on $R$ and $\theta$. Therefore 
the two-body interaction in the Hamiltonian is independent of 
the angles $\phi$ and $\psi$. The integrals of motion of the system may be 
constructed in terms of these new coordinates. The angular momentum 
operator is given by
\beq
L ~=~ \sum_i (x_i p_{y_i} - y_i p_{x_i}) ~=~ -i\frac{\partial}{\partial \psi} 
\eeq
which commutes with the Hamiltonian. There 
exists another constant of motion given by
\beq
Q ~=~ i ~[~ x_2 \frac{\partial}{\partial x_1} + y_2 \frac{\partial}{\partial 
y_1} - x_1 \frac{\partial}{\partial x_2} - y_1 \frac{\partial}{\partial 
y_2} ~] ~=~ -i\frac{\partial}{\partial \phi} ~.
\eeq
Since Q is antisymmetric, acting on a symmetric state produces an 
antisymmetric state and vice versa. We therefore refer to this as a 
supersymmetry operator (SUSY). The operator $Q$ is similar to the SUSY 
operator discovered in the many-anyon problem by Sen \cite{susy}. 
Note that 
the differential operator  for both angular momentum and the SUSY 
operators has a very simple form in the hyperspherical coordinates. The 
states can therefore be labeled by the quantum numbers associated with 
these two operators which we denote by $l$ and $q$ respectively. With 
SUSY, the two-body problem is integrable. (The four constants of motion
are the Hamiltonian $H$, the angular part of $H$, $L$ and $Q$). Note that we 
have $QX=0$ which makes calculations simple. It is easy to check that 
the bosonic ground state of the Hamiltonian has the quantum numbers $l$ and 
$q$ of the angular momentum and SUSY operators equal to zero. 

We would like to emphasize that the eigenstates of the SUSY operator $Q$ are 
neither symmetric nor antisymmetric, unless the eigenvalue $q=0$. After 
finding a simultaneous eigenstate of $H$, $L$ and $Q$, we can separate it into
symmetric (bosonic) and antisymmetric (fermionic) parts. These parts
are individually eigenstates of $Q^2$ but not of $Q$. Specifically, we have
$Q \Psi_B = q \Psi_F$ and $Q \Psi_F = q \Psi_B$, where $B$ and $F$ denote 
bosonic and fermionic states respectively. Then $\Psi_B \pm \Psi_F$ are 
eigenstates of $Q$, while $\Psi_B$ and $\Psi_F$ are eigenstates of $Q^2$.

The two-body Hamiltonian in terms of the hyperspherical coordinates is  
given by
\beq
H ~=~ -\frac{1}{2} ~[~ \frac{\partial^2}{\partial R^2} +\frac{3}{R} 
\frac{\partial}{\partial R} -\frac{\Lambda^2}{R^2} - R^2 ~] ~+~ g_1 
\frac{2}{R^2 \sin^2(2\theta)} ~, 
\label{twoham}
\eeq
where the operator $\Lambda^2$ is the Laplacian on the sphere $S^3$ and is 
given by
\beq
- \Lambda^2 ~=~ \frac{\partial^2}{\partial \theta^2} 
- \frac{2\sin(2\theta)}{\cos(2\theta)}\frac{\partial}{\partial \theta} 
+ \frac{1}{\cos^2(2\theta)} ~\Bigl[ ~\frac{\partial^2}{\partial \phi^2} 
+ 2\sin(2\theta) \frac{\partial^2}{\partial \phi\partial 
\psi} + \frac{\partial^2}{\partial \psi^2} ~\Bigr] ~.
\eeq

The interaction in the Hamiltonian is independent of the angles $\phi, 
\psi$ and depends only on $R,\theta$. The operators $L$ and $Q$ 
commute with the Hamiltonian since they commute with the noninteracting ($g=0$)
Hamiltonian. We thus label the states with the eigenvalues of these 
operators for all $g_1$. Each of these states is four-fold degenerate: 
Under parity, $L \rightarrow -L$  and $ Q \rightarrow Q$ and the 
Hamiltonian is invariant under parity. Therefore the  
states labeled by quantum numbers $(l,q)$ have the same energy as $(-l,q)$. 
The Hamiltonian is also invariant under the transformation $\vec 
r_1 \rightarrow -\vec r_1$ and $\vec r_2 \rightarrow \vec r_2$. This is a 
discrete symmetry peculiar to this system. 
Under this transformation $L \rightarrow L$  and $ Q \rightarrow -Q$.
Therefore the states labeled by quantum numbers $(l,q)$ have the same 
energy as $(l,-q)$. Combining the two we get the four-fold degeneracy of 
the states. Later we will find that the states with $(l,q)$ have the same 
energy as $(q,l)$ since interchanging $q$ and $l$ leaves the differential 
equation invariant; therefore the energy of these two states must be the same. 
We thus have an eight-fold degeneracy for the levels for which $|q|$ and $|l|$ 
are nonzero and different from each other. Note that this degeneracy is a 
subset of the degeneracy of the noninteracting system. If $|l|=|q|$ is 
nonzero, we have a four-fold degeneracy. Finally, there is a four-fold 
degeneracy between the states $(\pm l,0)$ and $(0, \pm l)$ if $l \ne 0$. 

\subsection{Solutions of the eigenvalue equation}

We are now interested in solving the eigenvalue equation given by
$ H\Psi = E\Psi$. Following the remarks made in the previous subsection, 
we may in general write
\beq
\Psi ~=~ F(R) ~\Phi(\theta, \phi, \psi) ~.
\eeq
The eigenvalue equation separates into angular and radial 
equations. The angular equation is given by
\beq
(~ \Lambda^2 + \frac{4g_1}{\sin^2(2\theta)} ~) ~\Phi ~=~ \beta (\beta + 2) ~
\Phi ~,
\label{angle}
\eeq
where $\beta \ge -1$, and the radial equation is given by 
\beq
\frac{d^2F}{dR^2} +\frac{3}{R} \frac{dF}{dR} +(~ 2E - R^2 - \frac{\beta 
(\beta +2)}{R^2} ~) ~F ~=~0 ~. 
\eeq
The radial equation can be easily solved using the methods outlined in 
the last section of \cite{LL}. The solution is given by
\beq
F(R) ~=~ R^{\beta} M(a,b,R^2) \exp{(-R^2/2)} ~,
\eeq
where $b=\beta+2 $ and $a=(\beta+2-E)/2$ and $M(a,b,R^2)$ is the confluent 
hypergeometric function. Demanding that $a = -n_r$ where $n_r$ is an integer, 
the energy is given by
\beq 
E ~=~ 2n_r + \beta +2 ~.
\label{eigenval}
\eeq
Note that $\beta$ is still unknown and has to be obtained by solving the 
angular equation. Nevertheless the tower structure of the eigenvalues 
built on radial excitation of the ground states is obvious from the above.  

The angular equation (\ref{angle}) may be solved with the ansatz
\beq
\Phi (\theta ,\phi ,\psi) ~=~ P(x) ~\exp (iq\phi) ~\exp (il\psi) ~,
\eeq
where $x =\sin(2\theta)$ and $l,q$ are the state labels in terms of the 
integer valued eigenvalues of the angular momentum and SUSY operators. The 
angular equation then reduces to a differential equation in a 
single variable $x$ for the function $P(x)$:
\beq
(1-x^2) \frac{d^2 P}{dx^2} 
-2x\frac{dP}{dx}-\frac{1}{4(1-x^2)} ~[~ q^2+2x q l+l^2 ~]P 
-\frac{g_1}{x^2}P+\frac{\beta (\beta +2)}{4}P ~=~ 0 ~.
\label{px}
\eeq
Note that the equation has four regular singularities at $x=0, 1, 
-1,\infty$ (the singularity at $\infty$ does not play any role since $x$ 
is bounded). Therefore the solution is of the form \beq
P(x) ~=~ |x|^a (1-x)^b(1+x)^c \Theta^{a,b,c}(x) ~.
\eeq
One can now fix $a,b,c$ to cancel the singularities. We find that
\beq
b ~=~ \frac{|l+q|}{4} \quad {\rm and} \quad c ~=~ \frac{|l-q|}{4} ~.
\eeq
Since $l$ and $q$ are integer valued, the values of $b$ and $c$ are 
restricted. The other exponent $a$ is given by
\beq
a(a-1)~=~g_1 ~=~ g(g-1) \quad {\rm with} \quad a ~\ge ~0 ~,
\label{a2g}
\eeq
where we have already defined $g_1$ through Eq. (\ref{g12}) in terms of $g$.
Note that we have used the symbol $a$ instead of $g$. As shown in the
Appendix, we have to take $a=g$ if $g \ge 1/2$. But if $g < 1/2$, we have to
generally consider a linear superposition of solutions with $a$ equal to $g$
and $1-g$ (more on this later).

We finally arrive at the required differential equation from which the 
eigenvalues are determined,
\bea 
(1-x^2)\frac{d^2\Theta}{d x^2} &+& 2[a/x -(b-c) -(a+b+c+1)x] ~\frac{d 
\Theta}{d x} \nonumber \\ 
&+& ~[~ \frac{(\beta +1)^2}{4} -(a+b+c+1/2)^2 +2a(c-b)/x ~] ~\Theta ~=~0 ~.  
\label{heun}
\eea
For $g=0$, the solutions are simply Jacobi polynomials and the full solution 
for the angular part is given in terms of the spherical harmonics on a 
four-dimensional sphere. In general, this differential equation is known as 
the Heun equation whose solutions $\Theta^{a,b,c}(x)$ are characterized by the 
so-called $P$-symbols \cite{bateman}. The Heun equation is exactly solvable if 
either $l$ or $q$ vanishes, i.e., if $b=c$ as discussed in the next 
subsection. The equation is also exactly solvable at an infinite number of 
isolated points in the space of parameters $(a,b,c)$. These are isolated 
points because if we vary $a$ slightly away from any one of them, the 
equation 
is not exactly solvable. Note that $b,c$ take discrete values and cannot be 
varied continuously. 

\subsection{Polynomial solutions}

Let us first consider a class of solutions which are polynomials in $x$. We 
may then write
\beq
\Theta(x) ~=~ \sum_{k=0}^p ~C_k x^k ~,
\eeq
where we may define $C_0 = 1$.
Substituting this in the differential equation for $\Theta$, we  
see that the $C_k's$ satisfy a three-term recursion relation given by
\bea
(k+2)(k+1+2a) ~C_{k+2} ~&-&~ 2(b-c)(k+1+a) ~C_{k+1} \nonumber \\
&+& ~[~ \frac{(\beta +1)^2}{4} -(a+b+c+k+1/2)^2 ~] ~C_k ~=~ 0 ~, 
\label{recursion}
\eea
which is in general difficult to solve. 
However there are two special cases when polynomial solutions are possible.
(i) For $b=c$, this reduces to a 
two-term recursion relation which can be easily solved to obtain all the 
energy levels. This is an example of a Conditionally Exactly 
Solvable (CES) problem \cite{D93} in which the full spectrum is exactly
solvable for some special condition (like $b=c$ here). (ii) The other case is 
when the coefficient of $C_k$ is zero with $k=p$ (where $p \ge 1$), i.e.
\beq
E ~=~ 2n_r+2a+2b+2c+2p+2 ~,
\label{E}
\eeq
in which case one has a polynomial solution of degree p. This is an example 
of a Quasi-Exactly Solvable (QES) problem when only a finite number of states
are exactly solvable for some given values of the parameters. As far as we 
are aware, this is the first example where both CES and QES solutions exist in 
the same problem. We now discuss both types of solutions in detail. 

\noindent
(i) CES-type Solutions:  To see the solutions explicitly, define $$y ~=~ 
x^2 ~.$$ In terms of the variable $y$ the differential equation is written as,
\beq
y(1-y)\frac{d^2\Theta}{d y^2} + 2 ~[~ a' - (b'+ c' 
+1)y ~]\frac{d \Theta}{d y} - b'c'\Theta ~=~ 0 ~,  
\eeq
where
\bea
a' & ~=~ &  a+1/2 ~, \nonumber \\
b' & ~=~ &\frac{1}{2}[a+2b+1+\frac{\beta}{2}] ~, \nonumber \\
c' & ~=~ &\frac{1}{2}[a+2b-\frac{\beta}{2}] ~. 
\eea
This is now a hypergeometric equation whose solutions are given by
$F(b',c',a'; y)$. Note that we need to concentrate only on the solutions for 
$0\le y \le 1$. The hypergeometric series terminates 
whenever $b'$ or $c'$ is a negative integer or zero. In our case $b'$ is 
always positive hence the solutions are given when $c' =-m$, where $m$ is 
an integer. Therefore $\beta =~ 2m+2a+4b$. The energy eigenvalues are obtained 
by substituting this in Eq. (\ref{eigenval}), 
\beq
E ~=~ 2n_r+2a+4b+2m+2 ~.
\eeq
Since $b=c$, we must have either $l=0$ or $q=0$. The energy varies linearly 
with $a$ as in the case of the exact solutions of the many-body problem. The
expression for $a$ in terms of $g$ will be clarified in the next subsection.
It might be tempting to conclude that all polynomial solutions vary linearly
with $g$. In fact that is not so as can be seen from the following examples.

\noindent
(ii) QES-type Solutions: We can have polynomial solutions of degree $p$ 
(where $p \ge 1$) if $b$, $c$ 
and $g$ satisfy some specific relations. Let us consider the case 
\beq
\Theta(x) ~=~ 1+\delta x ~.
\eeq
This is a solution if 
\beq
a ~=~ \frac{b+c}{(b-c)^2-1} -1
\label{constraint1}
\eeq
and $\delta = b-c$. We have implicitly assumed that $b$ is not equal to 
$c$ since otherwise this solution is trivial and of CES type. (Once again, 
$a$ can be equal to either $g$ or 
$1-g$ as discussed in the previous subsection). Then $\beta = 2a+2b+2c+2$,
and the full spectrum is given by
\beq
E ~=~ 2n_r+2a+2b+2c+4 ~.
\eeq
We should point out that a solution of this kind is only possible for fairly
large values of $l$ and $q$; the minimum values needed are $|l|=|q|=3$, in
which case $a=1/5$.

Similarly, there is a polynomial solution of the form
\beq
\Theta (x) = 1 + \delta x + \epsilon x^2 ~,
\eeq
if
\bea
a ~&=&~ 2 \alpha - \frac{3}{2} + {\sqrt {4 \alpha^2 - \alpha + \frac{1}{4}}} ~, 
\nonumber \\
{\rm where} \quad \alpha ~&=&~ \frac{b+c}{(b-c)^2 -4} ~.
\label{constraint2}
\eea
The spectrum is given by
\beq
E ~=~ 2n_r+2a+2b+2c+6 ~.
\eeq
These expressions for the energies are nonlinear in $a$  because of the 
constraints (\ref{constraint1}) or (\ref{constraint2}). We should however 
caution that these are isolated solutions since $b$ and $c$ can only take 
discrete values; hence the above solutions do not vary smoothly with $a$. 
In general, solutions similar to the above may be constructed for every degree 
$p$ of the polynomial; the corresponding energies are given by Eq. (\ref{E})
where $a$ is given by a function of $b$ and $c$ which can be derived by 
solving $p+1$ recursion relations obtained from Eqs. (\ref{recursion}) by
setting $k=-1,0,1,...,p-1$. 

\subsection{Numerical analysis}

We now consider the numerical solution of some low-lying states of the 
two-body problem since the polynomial solutions described in the previous 
subsection do not exhaust the full spectrum.

The noninteracting limit of the system is $g=0$ where we 
have the solutions corresponding to a four-dimensional oscillator. These are 
simply the spherical harmonics on a four-sphere $Y_{k,l,q}$, where  
$k=0,1,2,\cdots$ and $|l|,|q| \leq k$ label the states. When the 
degeneracy of these states is taken into account, all the states in the
noninteracting limit $g=0$ are completely specified. We now demand that the 
wave functions and energy levels should vary continuously with the parameter 
$g$ which is the interaction strength. We also require that the wave functions 
should not diverge at any value of $x$ in the interval $[-1,1]$.

For given values of $b \ne c$, we can numerically find the energy levels 
in two different ways. We can diagonalize the differential operator in 
(\ref{px}) in the basis of the noninteracting ($g=0$) states, or we can solve 
the differential equation (\ref{px}) or (\ref{heun}) directly for each 
state.  We have used both methods and will present the results below.

In order to proceed further, it is necessary 
to clarify the dependence of $a$ on $g$ in Eq. (\ref{a2g}). 
By the arguments given in the Appendix, $a=g$ if $g \ge 1/2$. If $g < 1/2$, 
$a$ can be equal to either $g$ or $1-g$; for any given state, we choose the 
value which is continuously connected to the noninteracting solution at $g=0$.
Namely, any solution of the above kind must, at $g=0$, go either as $1$ or 
as $x$ near $x=0$; we then choose $a=g$ or $a=1-g$ in the two cases 
respectively for $0 < g < 1/2$. In particular, an exact solution which, 
near $x=0$, goes as $1$ at $g=0$ will go as $|x|^g$ for all $g > 0$. An exact 
solution which goes as $x$ at $g=0$ will go as $|x|^{1-g}$ upto $g=1/2$ and 
then as $|x|^g$ for $g > 1/2$. As a result solutions of this form are 
discontinuous  at $g=1/2$. In general, when solving numerically for the
non-exact solutions, we have to allow a superposition of both $|x|^g$ and 
$|x|^{1-g}$. For $g \ge 1/2$, however, $a$ must be equal to $g$.

When solving the differential equations (\ref{px}) or 
(\ref{heun}), we have to consider the two 
regions separately. According to the rules discussed in the Appendix, for 
$0 \le g < 1/2$, we take the function $P(x)$ to go as 
\beq
P(x) ~=~ |x|^g ~+~ d ~{\rm sgn} (x) ~|x|^{1-g} ~,
\label{pxg}
\eeq
near $x=0$, and we vary both the coefficient $d$ as well as $\beta$ in Eq. 
(\ref{px}) till we find that the solution of (\ref{px}) does not diverge 
at $x = \pm 1$. Both $d$ and $\beta$ depend on $g$; it is possible to
determine the limiting values $d(0)$ and $d(1/2)$ as follows. At $g=0$,
suppose that the Jacobi polynomial is normalized so that $P=1$ and $P^{\prime}
=C_J$ at $x=0$. On the other hand, from Eq. (\ref{recursion}), we see that 
$C_1 = b-c$ for any nonzero $g$, no matter how small. By taking the limit 
$g \rightarrow 0$ in Eq. (\ref{pxg}), we therefore find that $d(0)=C_J -b+c$.
At the other end, $d(1/2)=\pm 1$ because, as we will show next, $P(x)$ must 
vanish for either $x \ge 0$ or for $x \le 0$ for all $g \ge 1/2$ (except for 
the exact polynomial solutions discussed previously). 

For $g \ge 1/2$, we must take the solution of the differential equation
to go either as (a) $P(x) = 0$ for $x \le 0$ and $\sim x^g$ 
for $x$ small and positive, or as (b) $P(x) = 0$ for $x \ge 0$ and $\sim 
(-x)^g$ for $x$ small and negative. In either case, we vary the energy till we
find that the solution of (\ref{heun}) (with $a=g$) does not diverge at 
$x=1$ and $-1$ for (a) and (b) respectively. It is interesting to note that for
each such solution, $P(x)$ vanishes identically in one of the half-intervals 
$[-1,0]$ or $[0,1]$. Note also that if $\Theta^{a,b,c} (x)$ is a solution
which is only nonzero for $x \ge 0$, then $\Theta^{a,c,b} (x)= \Theta^{a,b,c} 
(-x)$ will be a solution which is only nonzero for $x \le 0$; further, the two
solutions will have the same energy.

In contrast to the above method for solving the Heun equation (\ref{heun})
directly, the numerical diagonalization procedure for finding the eigenvalues
involves solving the eigenvalue equation (\ref{angle}). The basis for
diagonalization is provided by the eigenstates of the Laplacian 
$\Lambda^2$ on
$S^3$, namely the spherical harmonics on a four-sphere $Y_{k,l,q}$. For
non-zero interaction strengths, the singular interaction is handled by
multiplying the non-interacting eigenstates by $|x|^g$. The resulting basis is
nonorthogonal, and the diagonalization procedure is fairly straightforward
though cumbersome. We have truncated the basis such that the highest 
energy state in the basis is $100\hbar\omega$. The results for the low 
lying states in the  spectrum obtained through both
methods are displayed in Fig. 1 (for $0\leq g \leq 1$)and Fig. 2 (for $g\geq
1$). For $g < 1/2$, it is more convenient to use the diagonalization procedure
since the direct solution of Eq. (\ref{px}) requires one to numerically fix
two separate parameters $d$ and $\beta$. On the other hand, it is easier to
solve the Heun equation (\ref{heun}) for $g \ge 1/2$ since one has to fix only
one parameter $\beta$. In general we have used both methods to arrive at the 
spectrum of low-lying states. For small values of g ($<1$), however, the 
solutions 
of the differential equation produce eigenvalues which are somewhat smaller 
than the ones obtained by the diagonalization procedure. For large values of 
$g$, however, there is no perceptible difference between the results from the 
two methods. 

Fig. 1 shows the energies for some values of $l$ and $q$. Each level is
labeled by $(l,q;D)$, where $D$ is the degeneracy of the level away from 
the noninteracting limit; the degeneracy is computed by counting the allowed 
values of $\pm l$ and $\pm q$ using the parity and supersymmetry 
transformations for a given level. A subscript 'r' on the label $(l,q,D)$ 
denotes the radial excitation which is simply inferred from the existence of 
the towers. The bosonic ground state has 
the predicted behavior for all $g$; it is linear with a slope $2$ as a 
function of $g$. The corresponding wave function goes as $|x|^g$ as $x 
\rightarrow 0$. In contrast, the level $(0,0;1)$ starting at $E=4$ has 
an entirely different behavior . It is exactly
solvable for all $g$. According to the previous discussion, $dE/dg =-2$ for 
$g<1/2$ (with the wave function going as $|x|^{1-g}$ for small $x$) and 
$dE/dg =2 $ for $g>1/2$ (with the wave function going as $|x|^g$); thus
$dE/dg$ is discontinuous at $g=1/2$. 

At $g=0$, the slopes $dE/dg$ for all the levels can be calculated using
first-order perturbation theory as shown in the next subsection. For {\it
large} values $g$, we find that all the energies converge to $2g$ plus even 
integers as shown in Fig. 2. This amazing behavior can be understood using 
the WKB method as shown in subsection E. 

\subsection{Perturbation theory around $g=0$}

It is interesting to use perturbation theory to calculate the changes in 
energy from $g=0$, and to compare the results with the numerical analysis. We 
will only describe first order perturbation theory here, and the example we 
will consider is the fermionic ground state which is doubly degenerate for 
$N=2$. 

In general, naive perturbation theory fails at $g=0$ because most $g=0$ 
eigenstates do not vanish as $X_{ij} \rightarrow 0$; hence the expectation 
value of $1/X_{ij}^2$ diverges. This problem can be tackled by using a special 
kind of perturbation theory first devised for anyons \cite{pert}. We will 
first describe the idea for $N$ particles and then specialize to $N=2$.
Instead of solving the equation
$H \Psi = E \Psi$, where H is given in Eq. (\ref{ham2}), we perform a 
similarity transformation to ${\tilde H} = X_N^{-g} H X_N^g$ and ${\tilde \Psi} 
= X_N^{-g} \Psi$, where $$X_N \equiv \prod_{i<j} |X_{ij} |^g ~.$$ We then 
find that ${\tilde H} = H_0 + {\tilde V}$, where
$H_0$ is the noninteracting Hamiltonian (with $g_1=g_2=0$), and
\beq
{\tilde V} ~=~ g \sum_{i \ne j} ~\frac{1}{X_{ij}} ~(x_j 
\frac{\partial}{\partial y_i} - y_j \frac{\partial}{\partial x_i}) ~.
\eeq
The first order changes in energy may now be obtained by calculating the
expectation values (or matrix elements, in the case of degenerate states) of 
$\tilde V$ in the zeroth order (noninteracting) eigenstates. These expectation
values can be shown to be convergent for all states. Note that $\tilde V$ only 
contains two-body terms. Although $\tilde V$ is not hermitian, it is guaranteed 
that its expectation values are real because the original problem has a 
hermitian Hamiltonian $H$.

For $N=2$, we find that
\beq
{\tilde V} ~=~ -2g ~(~ \frac{1}{R} \frac{\partial}{\partial R} + \frac{\cot 2
\theta}{R^2} \frac {\partial}{\partial \theta} ~) 
\label{pertv}
\eeq
in hyperspherical coordinates.
Note that $\tilde V$ commutes with both $L$ and $Q$, so that we only have to
consider its matrix elements within a particular block labeled by the
eigenvalues $l$ and $q$. Let us now use (\ref{pertv}) to compute the first 
order change in the states which have $E=3$ at $g=0$. There are four such
states, with $l=\pm 1$ and $q= \pm 1$ (labeled $(1,1;4)$ in Fig. 1); two of 
these states are actually the 
ground states of the two-fermion system. Due to parity and SUSY, these four
states remain degenerate for all $g$. Hence it is sufficient to calculate the
first order change in the state with, say, $(l,q)=(1,1)$. Since this state is
unique at $g=0$, we only need to do non-degenerate perturbation theory with 
${\tilde V}$. The normalized wave function for this state is
\beq
\Psi ~=~ \frac{1}{\pi {\sqrt 2}} ~(\cos \theta - \sin \theta) ~\exp [i(\phi + 
\psi)] ~R \exp [-R^2 /2] ~.
\eeq
We now obtain the expectation value 
\beq
\int_0^{\infty} R^3 ~dR ~\int_{-\pi/4}^{\pi/4} \cos (2\theta) ~d\theta ~
\int_{-\pi/2}^{\pi/2} d\phi ~\int_{-\pi}^{\pi} d\psi ~\Psi^* ~{\tilde V}
\Psi ~=~ g ~.
\eeq
We can see from Fig. 1 that this gives the correct first order expression
for the energy $E=3+g$ near $g=0$ for the states labeled $(1,1;4)$; their 
first radial excitations $(1,1;4)_r$ therefore have $E=5+g$.

We can similarly calculate the first order expressions for the energies 
near $g=0$ for 
all the other levels shown in Fig. 1. We find that $E=2+2g$ for the lower state 
labeled $(0,0;1)$ (i.e. the bosonic ground state); $E=4+2g$ for its 
radial excitation $(0,0;1)_r$, and the states $(2,0;4)$; $E=4-2g$ for the 
upper state $(0,0;1)$; $E=4$ for the states $(2,2;4)$; $E=5-g/2$ for the 
states $(1,1;4)$ and $(3,3;4)$; and $E=5+3g/2$ for the states $(3,1;8)$.

\subsection{Large-$g$ perturbation theory}

We can study the solutions of Eq. (\ref{heun}) for large values of $g$ by using
an expansion in $1/g$. For any value of $b$ and $c$, we will 
only study the lowest energy $E$, and we will calculate the leading order 
terms in $E$ and the wave function $\Theta (x)$. We first note that the terms 
of order $g^2$ in (\ref{heun}) can be satisfied only if $E=2g +O(1)$. Next, we 
assume that $E$ and $\Theta$ have WKB expansions \cite{wkb} of the form
\bea
E ~&=&~ 2g ~+~ 2b ~+~ 2c ~+~ 2 ~+~ f_0 ~+~ \frac{f_1}{g} ~+~ O(1/g^2) ~, 
\nonumber \\
\Theta ~&=&~ \exp [ ~w_0 (x) ~+~ \frac{w_1(x)}{g} ~] ~.
\eea
The boundary condition $\Theta =1$ implies that $w_0 (0) = w_1 (0) =0$.
To order $g$, Eq. (\ref{heun}) gives the first order differential equation 
\beq
(1 - x^2) ~\frac{d w_0}{dx} ~=~ b-c-\frac{f_0}{2} x ~.
\eeq
We now look at solutions which are
nonzero only for $x \ge 0$. We demand that $\Theta$
should neither diverge nor vanish (since the lowest energy solution should
be node less) anywhere in the range $0 \le x \le 1$. Hence the functions $w_0$
and $w_1$ should not diverge to $\infty$ or $-\infty$ in that range. This 
fixes $f_0 =2(b-c)$, so that
\bea
E ~&=&~ 2g ~+~ 4b ~+~ 2 ~=~ 2g ~+~ |l+q| ~+~ 2 ~, \nonumber \\
{\rm and} \quad \Theta ~&=&~ (1 ~+~ x)^{b-c} ~.
\label{wkb1}
\eea
Similarly, there are solutions which are nonzero only for $x \le 0$. These have
\bea
E ~&=&~ 2g ~+~ 4c ~+~ 2 ~=~ 2g ~+~ |l-q| ~+~ 2 ~, \nonumber \\
{\rm and} \quad \Theta ~&=&~ (1 ~-~ x)^{c-b} ~.
\label{wkb2}
\eea
We now go to order $1$ in Eq. (\ref{heun}). For solutions which are nonzero
only for $x \ge 0$, we find that
\bea
E ~&=&~ 2g ~+~ 4b ~+~ 2 ~+~ \frac{c^2 - b^2}{g} ~=~ 2g ~+~ |l+q| ~+~ 2 ~-~
\frac{lq}{4g} ~, \nonumber \\
{\rm and} \quad \Theta ~&=&~ (1 ~+~ x)^{b-c} ~\exp [~ \frac{(b-c)(b+c+2)}{2g} ~
( \ln (1+x) ~-~ \frac{x}{1+x} ) ~] ~.
\label{wkb3}
\eea
We can similarly find solutions which are nonzero only for $x \le 0$, by 
changing $x \rightarrow -x$ and interchanging $b \leftrightarrow c$, i.e. 
$l+q \rightarrow l-q$ and $lq \rightarrow -lq$, in Eq. (\ref{wkb3}).

We see from Fig. 2 that these formulae correctly describe the leading 
behavior of $E$. In fact, the large-$g$ behavior is already visible in Fig.
1, for some states,  as we approach $g=1$. The various levels shown in that 
figure have the
following WKB energies; $E=2g+2$ for both the $(0,0;1)$ states (one of these
is the bosonic ground state and the other is the fermionic ground state for
$g>1/2$ as discussed later); $E=2g+2+1/4g$ for the states $(1,1;4)$; 
$E=2g+2+1/g$ for the states
$(2,2;4)$; $E=2g+2+9/4g$ for the states $(3,3;4)$; $E=2g+4-1/4g$ for the
states $(1,1;4)$; $E=2g+4$ for the radial excitation $(0,0;1)_r$ and the 
states $(2,0;4)$; $E=2g+4+1/4g$ for the radial excitations $(1,1;4)_r$; and $E=
2g+4+3/4g$ for the states $(3,1;8)$. We have also checked that the leading 
order wave functions in Eqs. (\ref{wkb3}) agree remarkably well with the 
correct wave functions $\Theta (x)$ obtained by solving the Heun equation 
(\ref{heun}) even if $g$ is not very large.

It is easy to see from Eqs. (\ref{wkb1}-\ref{wkb2}) that for large $g$, the
ground state and also the excited states become infinitely degenerate. This is
so because one can choose the quantum numbers $l$ and $q$ in infinitely many
ways such that the energies are the same as $g \rightarrow \infty$. Further,
the spacings now become twice the spacing at $g=0$ since $l$ and $q$ have the
same parity mod $2$.

The large-$g$ behavior therefore displays a remarkable similarity to the
problem of a particle in an uniform magnetic field where the Landau level
spacing is twice the cyclotron frequency, and each level is infinitely 
degenerate.

\subsection{Fermionic Ground State Energy}

The fermionic ground state energy has a very unusual behavior as can be seen
from Fig. 1. For $0 < g < 0.367$, the ground state energy 
monotonically and nonlinearly increases from $3$ to $3.266$ along
the curve labeled $(1,1;4)$. Beyond this point, for $0.367 < g <0.5$,
the ground state energy monotonically and linearly decreases from $3.266$ to 
$3$ along the upper curve $(0,0;1)$ satisfying $E=4-2g$. For $g \ge 1/2$, the 
fermionic and bosonic ground state energies are identical and are given by the 
curve $(0,0;1)$ which satisfies $E_0 = 2 + 2g$, i.e., both the ground states
monotonically increase with $g$. Thus the fermionic
ground state consists of three pieces as a function of $g$, while the bosonic
ground state is given by the single line $E_0 = 2 + 2g$ for all $g \ge 0$.

For two particles, one can understand why the fermionic and bosonic ground 
state energies are identical for $g > 1/2$ as follows. In this range of $g$, 
the ultraviolet potential near $x=0$ is infinite and it prevents tunneling 
between the regions $x>0$ and $x<0$ (see the Appendix). For two identical 
particles, an exchange necessarily takes us from a region with $x>0$ to 
a region with $x<0$ according to (\ref{exchange}). If tunneling between the 
two regions is forbidden, it becomes impossible to compare the phases of the 
wave function of a given configuration of the two particles and the wave
function of the exchanged configuration. Thus it is impossible to distinguish
bosons from fermions if $g>1/2$, and their energy levels must be identical.

It is possible that the same argument will go through for more than two
particles; however we need to understand the ultraviolet regularization of the
three-body interactions properly in order to prove that rigorously. If the 
argument holds, then we would have the interesting result that the 
$N$-fermion ground state energy is also given by (\ref{egs}) for $g>1/2$,
while it may show one or more level crossings for $g<1/2$. 

\section{Discussion and Summary}

To summarize, we have studied a two-dimensional Hamiltonian whose eigenstates
have a novel two-particle correlation. We have shown the existence of several
classes of exact solutions in the many-body problem. We have analyzed the
two-particle problem in detail and shown that it is completely solvable by 
reducing it to an ordinary differential equation in one variable which can be
solved exactly for a subset of states and numerically otherwise. The 
two-body problem is integrable since there are four 
constants of motion in involution. We have also discussed perturbation theory
for both small and large coupling strengths. In the strong interaction limit,
the system simplifies and bears a remarkable resemblance to the Landau level
structure. 

We have also clarified in the Appendix the ultraviolet prescription 
which is required to make sense of an inverse-square (singular) potential 
especially at small coupling strengths. In 
particular, we emphasize that it is in general not sufficient to specify that 
the wave functions are regular and square integrable to obtain an energy 
spectrum uniquely when dealing with singular interactions. In some domains of 
the coupling strength, we also need to specify the ultraviolet regularization 
to make complete sense of the results. We do this by demanding that as the 
parameter $g \rightarrow 0$, the energy levels should smoothly approach the 
known noninteracting levels. We believe
that this discussion is quite general and may have a wider applicability
to Hamiltonians with singular interactions.

Interesting problems for the future would be to extend this analysis to more
than two particles, and to find an application of our model to some physical
system. Recently, we have come to know that our model has been generalized to 
three (and higher) dimensions with novel three-body 
(and many-body) correlations \cite{G96}.

Three of us (RKB, JL and MVNM) would like to acknowledge financial support
from NSERC (Canada). RKB would like to acknowledge the hospitality at the
Institute of Physics, Bhubaneswar where part of this work was done.
MVNM thanks the Department of Physics and Astronomy, McMaster University 
for hospitality.

\vskip 1 true cm
\centerline{\bf Appendix}

We begin directly from Eq. (\ref{px}). Given the real number $g \ge 0$ 
satisfying
$g_1 = g(g-1)$, $P(x)$ could go, as $|x| \rightarrow 0$, as either $|x|^g$ 
or $|x|^{1-g}$ or even as a general superposition of the two powers.
We therefore need to define the problem more carefully in order to pick out 
a desired solution \cite{S58}. 

As mentioned above in the text, we demand the following. Firstly, the limit 
$g=0$ should give all the noninteracting two-particle solutions, both bosonic 
and fermionic. Secondly, all the wave functions and energies $E$ should be 
{\it continuous} functions of $g$, but the
first derivative of $E$ need not be continuous (indeed $dE/dg$ is not always
continuous at $g=1/2$ as we saw earlier). Finally, for $g > 1$, the wave 
function should go as $|x|^g$, and not as $|x|^{1-g}$ which diverges at $x=0$.

>From these three requirements, it is clear that for $g \ge 1/2$, the wave
functions must go purely as $|x|^g$, whereas for $g < 1/2$, the wave function
could go either as $|x|^g$ or $|x|^{1-g}$ or a superposition of the two.

We will now show that we can satisfy the above requirements if we redefine the 
problem with a different potential
in an {\it ultraviolet} region $|x| < x_o$. We take the potential to be
\bea
V(x) ~&=&~ \frac{g(g-1)}{x^2} \quad {\rm for} \quad |x| > x_o ~, 
\nonumber \\
&=&~ \frac{u^2}{x_o^2} \quad {\rm for} \quad |x| < x_o ~, \nonumber \\
{\rm where} \quad u ~\tanh u ~&=&~ g ~\quad {\rm if} \quad 0 \le 
g < 1/2~, \nonumber \\
{\rm and} \quad u ~&=&~ \infty ~\quad {\rm if} \quad g \ge 1/2 ~.
\label{vx}
\eea
Eventually, of course, we have to take the limit $x_o \rightarrow 0$ to recover
our original problem. Note that the potential in the ultraviolet region is
not symmetric under $g \rightarrow 1-g$ for $g \le 1$. Hence the energy 
spectrum does not have this symmetry.

To see why Eqs. (\ref{vx}) work, we note that the wave function, for $|x|$ 
slightly greater than $x_o$ (where $x_o$ is much smaller than any physical
length scales like the width of the harmonic oscillator potential), is 
generally given by 
\bea
P(x) ~&=&~ x^g ~+~ d_+ ~x^{1-g} \quad {\rm if} \quad x > x_o ~, \nonumber \\
{\rm and} \quad P(x) ~&=&~ (-x)^g ~+~ d_- ~(-x)^{1-g} \quad {\rm if} \quad 
x < - x_o ~. 
\eea
(For the exceptional case $g=1/2$, we have to replace $|x|^g$ and $|x|^{1-g}$
by $|x|^{1/2}$ and $|x|^{1/2} \ln |x|$ respectively).

Now consider the first case in (\ref{vx}), i.e., $0 \le g < 1/2$. Since the 
energy $E$ is much less than the potential in the inside region $|x| < x_o$
(this is necessarily true for any finite value of $E$ as $x_o \rightarrow 0$),
the wave function in that region is given by
\beq
P(x) ~\simeq ~ \cosh ~[~ (~{u \over x_o} + O(x_o) ~)~ x ~+~ \delta ~]~,
\eeq
where $\delta$ can be a complex number, and the term of $O(x_o)$ arises from 
the energy $E$ which is much less than $(u /x_o)^2$. We now match the 
wave function and its first derivative, or, more simply, the ratio $P^{\prime}
(x)/P(x)$ at $x= x_o \pm \epsilon$ and at $x = -x_o \pm \epsilon$, where
$\epsilon$ is an infinitesimal number. We then find three possibilities.

\noindent
(i) The wave function may be even about $x=0$. Then $\delta =0$, and $d_+ = 
d_-$ must vanish as $x_o^{1+2g}$ as $x_o \rightarrow 0$. (The behavior of
$d_{\pm}$ can be deduced by equating the terms of $O(x_o^{-1})$ and $O(x_o)$ 
in $P^{\prime} /P$ at $x=x_o \pm$). In the limit $x_o \rightarrow 0$, therefore,
the wave function goes purely as $|x|^g$.

\noindent
(ii) The wave function may be odd about $x=0$. Then $\delta = i 
\pi /2$, and $d_+ = d_-$ must 
diverge as $x_o^{2g-1}$ as $x_o \rightarrow 0$. The wave function is
proportional to ${\rm sgn} (x) ~|x|^{1-g}$ in that limit.

\noindent
(iii) In the general asymmetric case, we find that we must have $\delta$ of
order $x_o^{1-2g}$, and $d_+ = - d_- = d$ of O(1). (This is found by equating 
terms of $O(x_o^{-1})$ and $O(x_o^{-2g})$ in $P^{\prime} /P$ at $x=\pm x_o$). 
The wave function is therefore a superposition of the form
\beq
P(x) ~=~ |x|^g ~+~ d ~ {\rm sgn} (x) ~|x|^{1-g} ~.
\eeq

The cases (i) and (ii) arise if either $l$ or $q$ is zero in Eq. (\ref{px}),
since the equation is invariant under $x \rightarrow -x$ in that case. This
is precisely when $b=c$ and the equation is exactly solvable. We thus see
that the even solutions go as $|x|^g$, while the odd solutions go as 
$|x|^{1-g}$. If neither $l$ nor $q$ is zero, i.e. $b \ne c$, we have case 
(iii) where a superposition of the two powers are required.

The second case in (\ref{vx}), i.e. $g \ge 1/2$, is relatively simpler to
analyze since the wave function must be zero in the inside region $|x| \le 
x_o$.  On imposing this condition on the wave function in the outside region, we
see that both $d_+$ and $d_-$ must vanish as $x_o \rightarrow 0$. Hence the 
wave function will go purely as $|x|^g$ in that limit. However, since there
is no tunneling possible through the infinite barrier separating $x > x_o$
from $x < - x_o$, we will generally have wave functions which are nonzero 
only for $x > x_o$ or only for $x < -x_o$. This is indeed true as we saw 
earlier for the solutions of the Heun equation for $g > 1/2$.

We would like to emphasize that the relation $u \tanh u = g$ in
Eq. (\ref{vx}) is absolutely essential in order to have the possibility of
$P(x) \sim |x|^g$ for $g < 1/2$. If $u$ were to take any other value,
we would find that $P(x)$ necessarily goes as $|x|^{1-g}$ in the limit $x_o 
\rightarrow 0$. A similar fine tuning of $u$ is necessary in the CSM for $g < 
1/2$.  Incidentally, the strongly repulsive potential in the
ultraviolet region explains the peculiar result that the bosonic ground state
energy increases monotonically with $g$ even though the potential away from
the ultraviolet region becomes more and more attractive as $g$ goes from $0$
to $1/2$. One can show from Eq. (\ref{vx}) that the integrated potential
$\int_{-1}^11 dx V(x)$ is actually {\it positive} and large if $x_o$ is 
small, and it increases as $g$ varies from $0$ to $1/2$.

Several comments are in order at this stage.

\noindent
(i) A similar fine tuning is also required in the CSM if $g <1/2$ is to be
allowed as has been done by several people \cite{MS94}.
Historically, both Calogero \cite{C69} and Sutherland \cite{S71} restricted 
themselves to $g>1/2$. In a sense they could do that since the free fermion
limit corresponds to $g = 1$; thereby they avoided the problem with $g<1/2$.
However, one cannot reach the free bosonic limit smoothly in that case. 
Both these authors believed $g<1/2$ to be unphysical because they chose a 
particular regularization. What we have argued here 
is that one can choose an alternative regularization (called the resonance
condition by Sutherland) which allows one to go continuously all the way upto 
$g=0$ and hence reach the free bosonic limit continuously.
We have not seen this being clearly stated in 
the CSM literature before, although Scarf \cite{S58} discusses this issue in
a different problem containing the inverse-square potential.

It is worth noting that CSM has only two-body interactions. Henca  the entire
discussion here is also valid in the many-body case in CSM. 
This is in  
contrast to our problem where, for $N>2$, one also has to analyze the 
ultraviolet regularization of the three-body interactions.
 
\noindent
(ii) One important consequence of our regularization is that many of the 
states have discontinuities in $dE/dg$ at $g=1/2$; the fermionic ground state 
also has a level crossing at $g=0.367$. Further, for each value 
of $g_1 <0$, there are two possible ground states since the ultraviolet 
regularization depends on $g$ and not on $g_1$. 

\noindent
(iii) Put differently, our Hamiltonian $H$ has several self-adjoint extensions 
(SAE) for each value of $g_1$. What we have done is to choose a particular 
SAE for $g_1>0$ and 
two different SAE for $g_1<0$. As a result, we have found that for every 
value of $g_1$ in the range $-1/4<g_1<0$, there are two possible ground state 
energies since, as seen above, the SAE depends on $g$ rather than on $g_1$. 
Actually, there is an even more general SAE possible for any value of
$g_1$ where another real parameter (besides $g$) has to be introduced; however
we shall not discuss that here.

\begin{figure}
\caption{Spectrum of some low energy states in the two-body problem as a 
function of the interaction parameter $g$. The states are labeled by 
$(l,q;D)$, where the first two entries denote the two angular quantum numbers 
and the last entry shows the degeneracy of the level. Further the radial 
excitations are denoted by the subscript 'r'.}
\end{figure}

\begin{figure}
\caption{Spectrum of states as a function of $g$ showing the strong coupling 
behavior. We have shown $E-2g$ instead of the energy itself.}
\end{figure}

\end{document}